\documentclass[aps,prl,10pt,twocolumn,superscriptaddress]{revtex4}

\usepackage{graphicx}
\usepackage{amssymb, amsmath}
\usepackage{color}
\usepackage{ulem}
\usepackage{graphicx}
\usepackage{subfigure}
\usepackage{dcolumn}
\def\I{\uppercase\expandafter{\romannumeral 1}}
\def\II{\uppercase\expandafter{\romannumeral 2}}
\def\III{{\uppercase\expandafter{\romannumeral 3}}}
\def\IV{{\uppercase\expandafter{\romannumeral 4}}}
\def\V{{\uppercase\expandafter{\romannumeral 5}}}
\def\VI{{\uppercase\expandafter{\romannumeral 6}}}
\def\VII{{\uppercase\expandafter{\romannumeral 7}}}

\def\nn{\nonumber\\}

\def\nn{\nonumber\\}

\begin{document}

\title{Floquet engineering of multi-orbital Mott insulators: applications to orthorhombic titanates}

\author{Jianpeng Liu}
\affiliation{Kavli Institute for Theoretical Physics, University of California, Santa Barbara
CA 93106, USA}

\author{Kasra Hejazi}
\affiliation{ Department of Physics, University of California, Santa Barbara
CA 93106, USA}

\author{Leon Balents}
\affiliation{Kavli Institute for Theoretical Physics, University of California, Santa Barbara
CA 93106, USA}

\begin{abstract}
  We consider driving multi-orbital Mott insulators using laser
  radiation. We derive general expressions for periodically driven
  spin-orbital models using time-dependent perturbation theory in the
  strong interaction limit. We
  show that the effective exchange interactions of the Floquet
  spin-orbital Hamiltonian are highly tunable via variations of the
  frequency, amplitude, and polarization of the laser. We also take
  the effect of finite bandwidth of excitations into account and study
  possible heating effects. We further apply our formalism to
  orthorhombic titanates YTiO$_3$ and LaTiO$_3$ based on
  first-principles calculations, and find that the spin exchange
  interactions in these compounds can be engineered to a large extent
  by tuning the frequency and electric-field amplitude of the laser.
\end{abstract}

\maketitle

Periodically driven quantum systems have received significant
attention in recent years.  The typical theoretical prescription is to
use the Floquet formalism \cite{floquet-pr65, floquet-pra73}, which
allows for the description of a time-periodic system using some
effectively time-independent Hamiltonian dubbed as the ``Floquet
Hamiltonian", $H_{\textrm{F}}=i\hbar\log{U(T,0)}/T$, where $U(T,0)$ is
the time-evolution operator from time 0 to a full period $T$
\cite{bukov-aip-15}. 
Despite the problem of thermalization at long times \cite{floquet-infiniteT-prx14, moessner-pre14}, it has been argued that at experimentally accessible 
finite time scales the time evolution of the system is well described by the time-independent Floquet Hamiltonian \cite{saito-floquet-16}. 

Since the details of the Floquet Hamiltonian are crucially dependent
on the frequency, amplitude and polarization of the external drive,
the physical properties of a quantum system may be engineered using laser radiation.
Such ``Floquet engineering" has been extensively studied in the context of both single-particle \cite{takashi-aoki-prb09, floquet-transport-prb11, floquet-ti-np11, tanaka-prl10, floquet-silicene-prl13, platero-prl13, floquet-classification-prb11, levin-prx13,lindner-prx16, floquet-ti-surface-exp, floquet-volkov-np16} 
and many-body \cite{floquet-cuprate-science11,
  floquet-cuprate-nature14, floquet-k3c60-nature16,
  floquet-pairing-prb16, floquet-pairing-arxiv17, floquet-cdw-prl13,
  floquet-sdw-nm12, floquet-exchange-nc15,mentink-review-2017,
  floquet-kondo-arxiv17, floquet-chiral-sl-arxiv16,
  floquet-fractional-chern-insulator-prl14} models.  

Here, we contemplate applications to the solid state, i.e. Mott
insulating transition metal oxides, for which the orbital degrees of
freedom plays an essential role\cite{tokura-orbitalreview, georges-review13, kugel82}.  We use many-body
time-dependent perturbation theory to derive general expressions for
effective spin-orbital model descriptions of multi-orbital Mott
insulators in the presence of laser irradiation.  We further include the
effects of the doublon-holon (DH) hopping, i.e.  the bandwidth of
excitations, into account in our perturbation theory
\cite{comment_second_paper}, which induces both real and imaginary
parts into the effective Floquet Hamiltonian projected onto
the spin-orbital subspace.  The
real part is interpreted as an effective spin-orbital model, and the
corresponding exchange interactions are renormalized by the periodic
driving, which allows for the Floquet engineering of the spin-orbital
states. The imaginary part on the other hand is related to the rate of
generation of DH pairs, and thus can capture the effects of
heating. 
We further apply our formalism to ferromagnetic YTiO$_3$ and
antiferromagnetic LaTiO$_3$ based on first-principles calculations. We
find that the antiferromagnetic and ferromagnetic Mott insulators
exhibit distinct responses to the laser radiation, and the exchange
interactions in these compounds can be engineered to a large extent by
moderate electric fields.

\textit{Floquet spin model:}
We start the discussion by reviewing the periodically driven Hubbard model:
\begin{equation}
H(t)=-\sum_{\langle ij\rangle \sigma} \left( t_{h}\,e^{i\,u_{ij}\,\sin\omega t} \ c_{i\sigma}^{\dagger} c^{\vphantom\dagger}_{j\sigma}
 + \text{h.c.} \right) + \ U\sum_{i}\hat{n}_{i\uparrow}\hat{n}_{i\downarrow}\;,
 \label{eq:hubbard}
\end{equation}
where $t_h$ is the hopping amplitude between sites $i$ and $j$, and
$U\!\gg\!t_h$ is the onsite Coulomb repulsion
energy. $u_{ij}\!=\!e\mathbf{E}_0\cdot\mathbf{r}_{ij}/\omega$, where
$\vert\mathbf{E}_0\vert$ denotes the magnitude of the AC electric
field with frequency $\omega$,
$\mathbf{E}(t)\!=\!\mathbf{E}_0\cos{\omega t}$, and
$\mathbf{r}_{ij}\!=\!\mathbf{r}_j-\mathbf{r}_i$ is the displacement
vector from lattice site $i$ to $j$.  The effective Floquet spin
Hamiltonian in such a periodically driven half-filled Hubbard model has
been extensively discussed in
Ref.~\onlinecite{floquet-exchange-nc15},~\onlinecite{floquet-chiral-sl-arxiv16},
and ~\onlinecite{sw-bukov-prl16}.  It has been shown that the
effective spin exchange interaction of the Floquet spin Hamiltonian
associated with the bond $\langle ij \rangle$ is renormalized due to
the periodic driving, and becomes dependent on both the frequency and
amplitude of the drive,
$J_{\langle ij
  \rangle}=\sum_{n=-\infty}^{\infty}4t_h^2\mathcal{J}_n^2(u_{ij})/(U-n\omega)$,
which includes contributions from all the virtual DH excitation processes which
absorb/emit $n$ photons weighted by $\mathcal{J}_n^2(u_{ij})$, where
$\mathcal{J}_n(u_{ij})$ is the $n$th Bessel function of the first
kind.  The energy of the virtually created DH pair which absorbs/emits
$n$ photons is just $U-n\omega$ if the effects of DH hopping are
neglected.

The Floquet spin model breaks down when the photon energy $\omega$
(setting $\hbar\!=\!1$) is in resonance with the interaction energy
$U$, i.e. $n\omega$ is around $U$. In such a resonance regime, the
periodic driving generates real DH pairs, and the description of
the system by the low-energy spin dynamics is no longer valid. The DH
excitation spectrum has a finite bandwidth $\sim\!4\sqrt{z-1}t_h$ ($z$
is the coordination number) due to hopping of the DH pairs. As a
result of this, real DH pairs are generated as long as the frequency
$n\omega$ ($n\!\sim\!\mathcal{O}(1)$) is within this excitation
band. On the contrary, when $n\omega$ is outside the DH band, the DH
creation rate is tiny and the description of the system by an
effective Floquet spin Hamiltonian is still valid, but the expression
for $J_{\langle ij \rangle}$ is modified by the DH hopping.

Following Ref.~\onlinecite{comment_second_paper}, a generic many-body state $|\Psi\rangle_t$ can be approximately expressed
as  $\vert\Psi\rangle_t\!\approx\!\vert\Psi_0\rangle_t+\vert\Psi_1\rangle_t$, where $\vert\Psi_n\rangle_t$ represents a state with $n$ doubly occupied sites (doublons), and the $n\!>1\!$ states have been neglected as they make higher order contributions
to the spin dynamics. The Schr\"odinger equation for the evolution of the two components of the state reads 
\begin{align}
&i \partial_t\vert\Psi_0\rangle_t =\hat{P}_0\,T_t\,\vert\Psi_1\rangle_t\;,\nn
&i \partial_t |\Psi_{1}\rangle_t =  U \ |\Psi_{1}\rangle_t + \ T_t |\Psi_0\rangle_t + \tilde{T}_t |\Psi_1\rangle_t \;,
\label{eq:schordinger}
\end{align}
where $T_t$ is the time-dependent hopping operator shown in
Eq.~(\ref{eq:hubbard}), $\hat{P}_n$ is the projector onto the subspace
with $n$ double occupancies, and
$\tilde{T}_t\!=\!\hat{P}_1 T_t \hat{P}_1$ is the hopping operator
projected onto the single-doublon space. Replacing $\tilde{T}_t$ by
its time average
$\overline{T}\!=\!(\omega/2\pi)\int_{0}^{2\pi/\omega} dt'
\tilde{T}_{t'}$, $\vert\Psi_1\rangle_t$ can be explicitly expressed as
a function of $\vert\Psi_0\rangle_t$ \cite{comment_second_paper}.
Plugging the expression of $\vert\Psi_1\rangle_t$ (in terms of
$\vert\Psi_0\rangle$) back into the first line of
Eq.~(\ref{eq:schordinger}), one would obtain the time-dependent
Sch\"ordinger equation projected onto the $\vert\Psi_0\rangle$
subspace. If we further assume that the dominant DH hopping processes
are those which create and annihilate the DH pairs at the same
lattices sites leaving the background spin configurations unchanged,
then it follows that \cite{comment_second_paper}
\begin{equation}
i\partial_t\vert\Psi_0\rangle_t=\sum_{\langle ij\rangle}\sum_{m,n=-\infty}^{\infty} H_{ij}^{mn}(t)\vert\Psi_0\rangle_t,
\label{eq:schordinger-spin}
\end{equation}
where
$H_{ij}^{mn}(t)=t_h^2 f_{jiij}^{mn}(t) \sum_{\sigma\sigma'}
c_{j\sigma'}^{\dagger} c^{\vphantom\dagger}_{i\sigma'} \, c_{i\sigma}^{\dagger}
c^{\vphantom\dagger}_{j\sigma}\ g_{dh}(n\omega)$, and
$g_{dh}(n\omega)\!=\!\langle \Psi_0\vert
c_{j\sigma}^{\dagger}c^{\vphantom\dagger}_{i\sigma}(U-n\omega + \bar{T})^{-1}
c_{i\sigma}^{\dagger}c^{\vphantom\dagger}_{j\sigma}\vert \Psi_0\rangle$ is the DH Green's
function, and
$f_{jiij}^{mn}(t)\!=\!-e^{i(m-n)\omega
  t}\mathcal{J}_{-n}(u_{ij})\mathcal{J}_m(u_{ji})$. We further assume
that the motions of the doublons and holons are uncorrelated, which
allows  $g_{dh}$ to be expressed as the convolution of the holon and doublon
Green's functions $g_d$ and $g_h$ \cite{comment_second_paper}. The
holon (doublon) Green's function $g_{h(d)}$ is then calculated using the
retraceable path approximation
$g_{h(d)}(E)=2(z-1)/(E (z-2)+z\sqrt{E^2-4(z-1)\bar{t}_h^2})$
\cite{brinkman-rice-prb70, comment_second_paper}, where
$\bar{t}_{h}\!=\!t_h\mathcal{J}_0(u_{ij})$ denotes the time-averaged
hopping amplitude.  In the regime $\omega\!\gg\!t_h^2/U$, the leading
order Floquet Hamiltonian is simply the time-average of the
right-hand-side of Eq.~(\ref{eq:schordinger-spin}).

\textit{Floquet spin-orbital model:}
The previous discussion of the periodically driven Hubbard model can be generalized to the case of
multi-orbital Mott insulators with local Kanamori interactions \cite{kanamori-63}
\begin{align}
H_{\textrm{K}}=&U\sum_{i,\alpha}\hat{n}_{i\alpha\uparrow}\hat{n}_{i\alpha\downarrow}+
U'\sum_{i,\alpha <\beta,\sigma,\sigma'}\hat{n}_{i\alpha\sigma}\hat{n}_{i\beta\sigma'}\;\nn
&-J_{\textrm{H}}\sum_{i,\alpha < \beta,\sigma,\sigma'}c^{\dagger}_{i\alpha\sigma}
c^{\vphantom\dagger}_{i\alpha\sigma'}c^{\dagger}_{i\beta\sigma'}c^{\vphantom\dagger}_{i\beta\sigma}\;\nn
&+J_{\textrm{P}}\sum_{i,\alpha < \beta,\sigma}
c^{\dagger}_{i\alpha\sigma}c^{\dagger}_{i\alpha-\sigma}c^{\vphantom\dagger}_{i\beta\sigma}c^{\vphantom\dagger}_{i\beta -\sigma}\;,
\label{eq:kanamori}
\end{align}
where $U$ and $U'$ are the intra-orbital and inter-orbital direct Coulomb interactions. 
$J_{\textrm{H}}$ and $J_{\textrm{P}}$ denote the Hunds' coupling 
and pair hoppings respectively; the sets of indices $\{i,j\}$,
$\{\alpha,\beta\}$, $\{\sigma,\sigma'\}$ denote, in turn, 
the lattice sites, orbitals and spin degrees of freedom.
As in the case of the Hubbard model, the effect of the periodic driving is manifested in the kinetic energy via the Peierls substitution,
\begin{equation}
T_t=\sum_{\langle ij \rangle, \alpha\beta,\sigma} \left( t_{ i \alpha, j\beta} \ e^{iu_{ij}\sin\omega\!t}c^{\dagger}_{i\alpha\sigma}c^{\vphantom\dagger}_{j\beta\sigma}+\textrm{h.c.} \right) ,
\label{eq:multi-hopping}
\end{equation}
where $t_{i\alpha,j\beta}$ represents the hopping amplitude from orbital $\beta$ at site $j$ to orbital $\alpha$
at site $i$. 

In the multi-orbital case, we also need to consider the crystal-field splittings ($H_{\textrm{CF}}$). 
In addition to the giant $t_{2g}-e_g$ splitting of typical perovskite transition-metal oxides, 
there may be additional splittings within
the $t_{2g}$ and/or $e_g$ manifold due to various distortions \cite{okanderson-prl04, kugel82}. 
Throughout this paper we only consider the $t_{2g}$ orbitals. Within the quasi-degenerate 
$t_{2g}$ levels we further include the crystal-field splittings,
\begin{equation}
H_{\textrm{CF}}=\sum_i\sum_{\alpha,\beta,\sigma}\epsilon_{i,\alpha\beta}\,c^{\dagger}_{i\alpha\sigma}c^{\vphantom\dagger}_{i\beta\sigma}\;,
\end{equation}
Including all these terms, we find the total periodically driven
Hamiltonian as $H_t=T_t + H_K + H_{\textrm{CF}}$
\footnote{The virtual excitations from the low-energy $3d$ levels to the high-energy $4p$ levels would induce
further splittings of the $3d$ states via the second-order Stark effect, but such splittings are very small
for $3d$ transition metal ions (see supplemental information).}

We consider the limit that the typical interaction energy
scale is much greater than the hopping energy scale 
and consider $T_t$ as a perturbation to
$H_{\textrm{K}}$. In the non-driven case, the low-energy physics is dominated by the spin and orbital dynamics, which is 
well described by the Kugel-Khomskii \cite{kugel-73,kugel82} and similar spin-orbital models, and can be derived
using second-order perturbation theory.
We generalize that approach to the case with periodic driving, and derive a time-dependent spin-orbital model using time-dependent perturbation theory. We consider the situation of one occupied electron at every site in the ground state of the static system, then make the assumption that $U'=U-J_{\textrm{H}}$ and $J_{\textrm{P}}\!=\!0$ \cite{georges-review13}. With such an assumption $H_{\textrm{K}}$ is rotationally invariant and there are only two distinct multiplet energy levels: $E_{\textrm{singlet}}\!=\!U, \hspace{12pt}$ for spin singlets, and $E_{\textrm{triplet}}\!=\!U-2J_{\textrm{H}}$ for spin triplets \cite{georges-review13}. Therefore, we expand an arbitrary many-body state $\vert \Psi\rangle_t$ as
$\vert\Psi\rangle_t\!\approx\!\vert\Psi_0\rangle_t + \vert\Psi_{1}^{\textrm{s}}\rangle_t+\vert\Psi_{1}^{\textrm{t}}\rangle_t$,
where $\vert\Psi_0\rangle_t$ represents the states without any double occupancy, and $\vert\Psi_{1}^{\textrm{s}}\rangle$ and $\vert\Psi_{1}^{\textrm{t}}\rangle$ denote the single-doublon states with spin singlets and triplets configurations.  As discussed above, we neglect the excited states with more than one doublons.

Time-dependent perturbation theory leads to the Schr\"odinger equation projected onto the zero-doublon subspace \cite{comment_second_paper},
\begin{equation}
i\partial_t \vert\Psi_0\rangle_t = \Big(\sum_{\langle ij\rangle,mn,a}
f^{mn}_{ij}(t) \ \hat{G}^{\textrm{a}}_{jiij}(n\omega)+H_{\textrm{CF}}\Big)\,\vert\Psi_0\rangle_t\;,
\label{eq:h-time-multi}
\end{equation}
where 
$f_{ij}^{mn}(t)=-e^{i(m-n)\omega t}\mathcal{J}_m(u_{ji})\mathcal{J}_{-n}(u_{ij})$,
$\hat{G}^{\textrm{a}}_{jiij}\!=\!\sum_{\alpha\beta\alpha'\beta',\sigma\sigma'}t_{i\alpha,j\beta}t_{j\beta',i\alpha'}
\,c^{\dagger}_{j\beta'\sigma'}c^{\vphantom\dagger}_{i\alpha'\sigma'}c^{\dagger}_{i\alpha\sigma}c^{\vphantom\dagger}_{j\beta\sigma}\,g_{dh}^{\textrm{a}}$,
and the superscript index  ``$\textrm{a}$" runs over $\{\textrm{s},\textrm{t}\}$.
$g_{dh}^{\textrm{s}}$ and $g_{dh}^{\textrm{t}}$ are the doublon-holon Green's functions in the spin singlet and triplet
configurations:
\begin{align}
&g_{dh}^{\textrm{s}}=\langle\Psi_0\vert c_{j\beta\sigma}^{\dagger}c^{\vphantom\dagger}_{i\alpha\sigma}\frac{\hat{P}_{1\textrm{s}}}{U-n\omega+\bar{T}^{\textrm{ss}}}
c_{i\alpha\sigma}^{\dagger}c^{\vphantom\dagger}_{j\beta\sigma}\vert\Psi_0\rangle\;,\nn
&g_{dh}^{\textrm{t}}=\langle\Psi_0\vert c_{j\beta\sigma}^{\dagger}c^{\vphantom\dagger}_{i\alpha\sigma}\frac{\hat{P}_{1\textrm{t}}}{U-2J_{\textrm{H}}-n\omega+\bar{T}^{\textrm{tt}}}
c_{i\alpha\sigma}^{\dagger}c^{\vphantom\dagger}_{j\beta\sigma}\vert\Psi_0\rangle\;.
\label{eq:multi-green-final}
\end{align}
We have made the following approximations in deriving Eq.~(\ref{eq:h-time-multi})-(\ref{eq:multi-green-final}).
First, we only consider the hopping processes
which create and annihilate DH pairs at the same sites, with a final \textit{spin-orbital} configuration which is identical to the initial configuration. Second we have neglected the doublon-holon hopping terms which convert a spin triplet to a singlet and vice versa. Lastly, we have time-averaged over the hopping operator projected onto the single doublon-holon space \cite{comment_second_paper}. 

In order to calculate the holon/doublon Green's function $g_h$/$g_h$ in the multi-orbital
case, we take the limit that the crystal field splitting (within the $t_{2g}$ or $e_g$ orbitals) 
is much larger than the intersite exchange energy, such that the occupied
orbital at site $i$ is uniquely determined and is denoted as $\vert 1\rangle_i$. 
In this classical-orbital regime, it is legitimate to introduce effective hoppings 
between the orbitals $\vert 1\rangle_i$
and $\vert 1\rangle_j$ for the singlet and triplet virtual excitations denoted as $t_{i1,j1}^{\textrm{s}}$ and 
$t_{i1,j1}^{\textrm{t}}$:
$(t^{\textrm{s}}_{i1,j1})^2=\sum_{\alpha}(\vert t_{i1,j\alpha}\vert^2+
\vert t_{j1,i\alpha}\vert ^2)/2$, and 
$(t^{\textrm{t}}_{i1,j1})^2=\sum_{\alpha\ne 1}(\vert t_{i1,j\alpha}\vert^2+
\vert t_{j1,i\alpha}\vert ^2)/2$.
Then the corresponding DH Green's functions $g_{dh}^{\textrm{s}}$
and $g_{dh}^{\textrm{t}}$ can be calculated using the single-orbital formalism discussed above.

When $\omega$ is much larger than typical exchange energies, the leading-order Floquet spin-orbital Hamiltonian is simply the time-average of the right-hand-side of Eq.~(\ref{eq:h-time-multi}).
For ${t}_{2g}$ orbitals 
the Floquet Hamiltonian can be rewritten in
terms of the $t_{2g}$ spin and orbital operators. After taking the expectation values of
the orbital operators, one obtains
\begin{align}
H_{\textrm{F}}^{\textrm{so}}=&\sum_{\langle ij\rangle, n}\Big(\,\mathcal{J}_n^2(u_{ij})(\gamma_1+\gamma_2)g_{dh}^{\textrm{s}}(n\omega)
(\mathbf{S}_i\cdot\mathbf{S}_j-\frac{1}{4})\;\nn
&-\mathcal{J}_n^2(u_{ij})(\gamma_1-\gamma_2)\,g_{dh}^{\textrm{t}}(n\omega)\,(\mathbf{S}_i\cdot\mathbf{S}_j+\frac{3}{4})\,\Big)\;.
\label{eq:h-floquet-t2g}
\end{align}
It follows that the effective spin exchange interaction 
associated with bond $\langle ij\rangle$ is
\begin{equation}
\bar{J}_{ij}=\sum_{n}\mathcal{J}_n^2(u_{ij})(\,(\gamma_1+\gamma_2)g_{dh}^{\textrm{s}}-(\gamma_1-\gamma_2)g_{dh}^{\textrm{t}}\,)\;.
\label{eq:floquet-t2g-exchange}
\end{equation}
where
$\gamma_1=\sum_{\alpha,\beta,\beta'=1}^{3}(t_{i\alpha,j\beta}\,t_{j\beta', i\alpha}\,\langle\hat{A}_{\beta'\beta}^{j}\rangle+i\leftrightarrow j)$,
and 
$\gamma_2=\sum_{\alpha,\alpha',\beta,\beta'=1}^{3}(t_{i\alpha,j\beta}\,t_{j\beta',i\alpha'}\,\langle\hat{A}^{j}_{\beta'\beta}\rangle\,\langle\hat{A}^{i}_{\alpha\alpha'}\rangle
+i\leftrightarrow j)$
and $\langle\hat{A}^{i}_{\alpha\alpha'}\rangle=\sum_{\sigma}\langle c^{\dagger}_{i\alpha\sigma}c_{i\alpha'\sigma}\rangle$ is the
expectation value of the orbital operator $\hat{A}^{i}_{\alpha\alpha'}$. The DH Green's function
in the singlet (triplet) configuration  $g_{dh}^{\textrm{s}(\textrm{t})}$ can be calculated using the single-orbital formalism in the regime of strong crystal-field splittings.

As in the case of non-driven system, the Floquet exchange interaction $\bar{J}_{ij}$  consists of two components: the antiferromagnetic component from all the singlet virtual excitations 
$\bar{J}_{ij}^{\textrm{AFM}}=\sum_n\mathcal{J}_n^2(u_{ij})(\gamma_2+\gamma_1)g_{dg}^{\textrm{s}}$, and the ferromagnetic component from all the triplet virtual excitations $\bar{J}_{ij}^{\textrm{FM}}=-\sum_n\mathcal{J}_n^2(u_{ij})(\gamma_1-\gamma_2)g_{dh}^{\textrm{t}}$.
Eq.~(\ref{eq:floquet-t2g-exchange}) suggests that the effective
exchange interactions in periodically driven multiorbital Mott insulators can be engineered by the periodic driving.

If the $\vert U-n\omega\vert$ and/or $\vert U-2J_{\textrm{H}}-n\omega\vert$ \cite{comment_nphoton}
is much greater than the typical hopping amplitudes, it is straightforward to show that $g_{dh}^{\textrm{s}}(n\omega)\!\approx\!
1/(U-n\omega)$ and $g_{dh}^{\textrm{t}}(n\omega)\!\approx\!1/(U-2J_{\textrm{H}}-n\omega)$. Eq.~(\ref{eq:floquet-t2g-exchange})
becomes
\begin{equation}
\bar{J}_{ij}=\sum_{n}\mathcal{J}_n^2(u_{ij})\left(\,\frac{\gamma_1+\gamma_2}{U-n\omega}-\frac{\gamma_1-\gamma_2}{U-2J_{\textrm{H}}-n\omega}\,\right)\;.
\label{eq:floquet-t2g-exchange-simple}
\end{equation}
In what follows we will show that ferromagnetic and antiferromagnetic
Mott insulators exhibit contrasting responses to laser radiation
due to the analytic structure of $\bar{J}_{ij}$ shown in Eq.~(\ref{eq:floquet-t2g-exchange-simple}).
On the other hand, if $\vert U-n\omega\vert\!<\!4\sqrt{z-1}t_h$ or $\vert U-2J_{\textrm{H}}-n\omega\vert\!<\!4\sqrt{z-1}t_h$, 
$g_{dh}^{\textrm{s}}$ or $g_{dh}^{\textrm{t}}$ has both real and imaginary parts. The non-vanishing imaginary part of the 
Floquet spin-orbital Hamiltonian
($\textrm{Im}[{H}_{\textrm{F}}^{\textrm{so}}]$) implies the norm of the
spin-orbital state $\vert\Psi_0\rangle$ decays with time, and the rate
of the DH generation is proportional to $\textrm{Im}[{H}_{\textrm{F}}^{\textrm{so}}]$.

\textit{Application to orthorhombic titanates:}
We apply the formalism discussed above to the orthorhombic 
perovskite titanates YTiO$_3$ and LaTiO$_3$. YTiO$_3$ is a ferromagnet Mott insulator with Curie temperature $T_{\textrm{C}}\!\approx\!27\,$K \cite{yto-spin-exp-prl02}, 
whereas LaTiO$_3$ is a ``$G-$type" antiferromagnetic Mott insulator (antiferromagnetic ordering in all the three spatial dimensions) with Neel temperature $T_{\textrm{N}}\!\approx\!146\,$K \cite{lto-distortion-prb03}. Both compounds can be considered as perovskite oxides with GdFeO$_3$-type distortions.
Moreover, there are other lattice distortions which split the otherwise degenerate $t_{2g}$ orbitals \cite{supp_info}. The crystal
field splittings $\sim\!0.1-0.4$\,eV
\cite{yto-orbitalorder-exp-jpsp01,titanates-wannier-dmft-njop05,solovyev-prb04},
are much larger
than the exchange energies. Hereafter we  assume that the orbital patterns are completely fixed by the crystal field splittings and neglect orbital fluctuations.

In order to evaluate the hopping parameters
(Eq.~(\ref{eq:multi-hopping})) and the crystal-field splittings
$H_{\textrm{CF}}$, we carried out bare density-functional-theory
(DFT)\cite{dft1,dft2} calculations with vanishing magnetizations for
LaTiO$_3$ and YTiO$_3$. The converged Bloch functions are then
projected onto the local $t_{2g}$ orbitals at the Ti sites to generate
Wannier functions with the $t_{2g}$ symmetry. Realistic tight-binding
models are then constructed in the basis of the $t_{2g}$ Wannier
functions \cite{MLWF-rmp, wannier90}. From these tight-binding models
we extract the hopping parameters and the crystal-field parameters
\cite{supp_info}. We have also estimated the Hubbard repulsion using
the linear-response method \cite{ldau-linear-response-prb05}. We find
that $U\!=\!3.83\,$eV, $J_{\textrm{H}}\!=\!0.64\,$eV in YTiO$_3$;
while $U\!=3.82\!$\,eV and $J_{\textrm{H}}=0.64\,$eV for
LaTiO$_3$. Using these parameters, the nearest-neighbor spin exchange within
the $ab$ plane $J_{ab}\!=\!-8.4$\,meV (minus sign means
ferromagnetic), and $J_c\!=\!-0.5\,$meV along the $c$ axis for
YTiO$_3$.  We note that the in-plane (inter-plane) ferromagnetic
exchange interaction of YTiO$_3$ is overestimated (underestimated) by
DFT.  On the other hand, for LaTiO$_3$ we get antiferromagnetic
exchange interactions with $J_{ab}\!=15.4\!$\,meV and
$J_c\!=11.9\!$\,meV, which are in good agreements with the spin-wave
measurements \cite{lto-spindynamics-prl00}.

\begin{figure}
\includegraphics[width=3.4in]{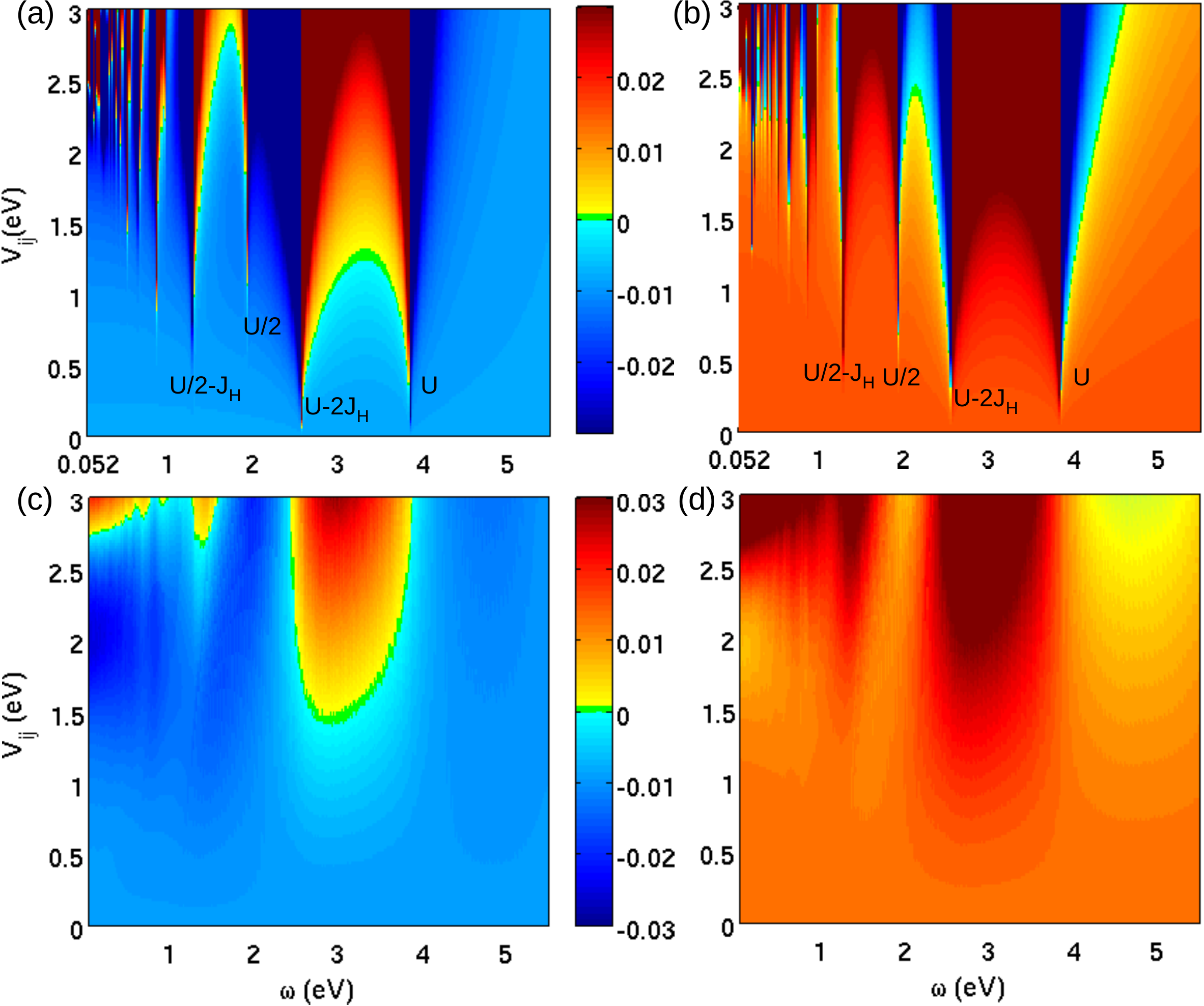}
\caption{The in-plane Floquet spin exchange interactions for YTiO$_3$ and LaTiO$_3$ as a function of the driving
frequency $\omega$ and the electric-field energy $V_{ij}$.
In (a)-(b), the bandwidth of the virtual doublon-holon excitations is neglected. In (c)-(d), the bandwidth
of the virtual excitation spectra  has been taken into account. (a) and (c) are for YTiO$_3$, and (b) and (d)
are for LaTiO$_3$.}
\label{fig:floquet-exchange}
\end{figure}

First we consider the Floquet exchange interactions neglecting the
bandwidth of the DH excitations as given by
Eq.~(\ref{eq:floquet-t2g-exchange-simple}).  Plugging all the hopping,
crystal-field, and interaction parameters evaluated from DFT into
Eq.~(\ref{eq:floquet-t2g-exchange-simple}), the effective spin
exchange interaction for the bond $\langle ij\rangle$ can be readily
obtained.  Neglecting the bandwidth of the DH excitations, in
Fig.~\ref{fig:floquet-exchange}(a)-(b) we plot the in-plane effective
spin exchange interactions for YTiO$_3$
(Fig.~\ref{fig:floquet-exchange}(a)) and LaTiO$_3$
(Fig.~\ref{fig:floquet-exchange}(b)) in the parameter space
$(\omega, V_{ij})$, where $V_{ij}\!=\!u_{ij}\omega$ is the
electric-field energy.

We notice that at relatively low frequencies and weak electric fields,
i.e. $u_{ij}\!\lessapprox\!1$, $\omega\!\ll\!U, U-2J_{\textrm{H}}$,
the effective ferromagnetic exchange interaction in YTiO$_3$ is
enhanced, but the antiferromagnetic exchange in LaTiO$_3$ is
suppressed as $V_{ij}$ increases.  Such opposite behaviors are
inherited from the analytic structure of
Eq.~(\ref{eq:floquet-t2g-exchange-simple}). When $u_{ij}$ is small and
$\omega\!\ll\!U, U-2J_{\textrm{H}}$, the dominant processes are those
with a small number of photon emissions/absorptions since the weight
$\mathcal{J}_{n}^2(u_{ij})\!\sim\!u_{ij}^{2n}$ for small $u_{ij}$.  Up
to second order in $u_{ij}$, the effective spin exchange interaction
(neglecting the effects of the virtual DH hopping) for bond
$\langle i j\rangle$ simplifies to
$\bar{J}_{\langle ij\rangle}\approx (\gamma_2+\gamma_1)(1/U+\delta
J_U)-(\gamma_1-\gamma_2)(1/(U-2J_{\textrm{H}})+\delta
J_{U-2J_{\textrm{H}}})$, where $\delta J_E$ (with
$E=U, U-2J_{\textrm{H}}$) is understood as the corrections to the
exchange interaction in the static limit,
$2\delta J_E=u_{ij}^2\,E\,(1/(E^2-\omega^2)-1/E^2)$.
We note that  $\delta J_E$ is always positive, and $\delta J_{U-2J_{\textrm{H}}}\!>\!\delta J_U$ for small $\omega$. 
As a result, when the total exchange is ferromagnetic (antiferromagnetic) in the static limit, the magnitude of the exchange is enhanced (diminished) as $V_{ij}$ increases. 

More interesting behavior appears when the frequency is on the same
order of magnitude as $U$ and is in between two virtual excitation
levels.  To be specific, if
$(U-2J_{\textrm{H}})/n_2\!<\!\omega<\!U/n_1$ ($n_1, n_2\!\sim\!1$), it
is convenient to express the photon energy as
$\omega\!=\!(U-2J_{\textrm{H}})/n_2+\delta\omega_2\!=\!U/n_1-\delta\omega_1$.
It follows that the dominant photon absorption/emission processes are
those with $n\!=\!0$, $n_1$ and $n_2$, and the effective Floquet
exchange interaction associated with bond $\langle ij\rangle$ is
approximated as
$J_{\langle ij
  \rangle}\!\approx\!\mathcal{J}_0^2(u_{ij})(\,(\gamma_2+\gamma_1)/U-(\gamma_1-\gamma_2)/(U-2J_{\textrm{H}})\,)+
\delta J_{\langle ij\rangle}^{n_1}+\delta J_{\langle
  ij\rangle}^{n_2}$, where
$\delta J_{\langle
  ij\rangle}^{n_1}\!=\!\mathcal{J}_{n_1}^2(u_{ij})(\,(\gamma_2+\gamma_1)/(n_1\delta\omega_1)+(\gamma_1-\gamma_2)/(2J_{\textrm{H}}-n_1\delta\omega_1)\,)$,
and
$\delta J_{\langle
  ij\rangle}^{n_2}\!=\!\mathcal{J}_{n_2}^2(u_{ij})(\,(\gamma_2+\gamma_1)/(2J_{\textrm{H}}-n_2\delta\omega_2)+
(\gamma_1-\gamma_2)/(n_2\delta\omega_2)\,)$. We note that
$\delta J_{\langle ij\rangle}^{n_1}$ and
$\delta J_{\langle ij\rangle}^{n_2}$ are always positive, which would
enhance the net antiferromagnetic exchange and suppress the net
ferromagnetic exchange.  This is clearly illustrated in
Fig.~(\ref{fig:floquet-exchange})(a)-(b) for
$U-2J_{\textrm{H}}\!<\!\omega\!<\!U$.  For YTiO$_3$
(Fig.~\ref{fig:floquet-exchange}(a)), the ferromagnetic exchange at
$V_{ij}\!=\!0$ is suppressed by turning on the electric field, and
becomes antiferromagnetic at some critical value of
$V_{ij}^{*}(\omega)\!\sim\!0.5-1.5\,$eV.  On the other hand, the
antiferromagnetic exchange for LaTiO$_3$ in
Fig.~(\ref{fig:floquet-exchange})(b) is enhanced as $V_{ij}$
increases.  Similarly, if $U/n_1\!<\omega<\!(U-2J_{\textrm{H}})/n_2$,
it is straightforward to show that periodic driving tends to
enhance the magnitude of the ferromagnetic exchange but suppress the
antiferromagnetic exchange, as shown in
Fig.~(\ref{fig:floquet-exchange})(a)-(b) for
$U/2\!<\!\omega\!<\!U-2J_{\textrm{H}}$.

\begin{figure}
\includegraphics[width=3.4in]{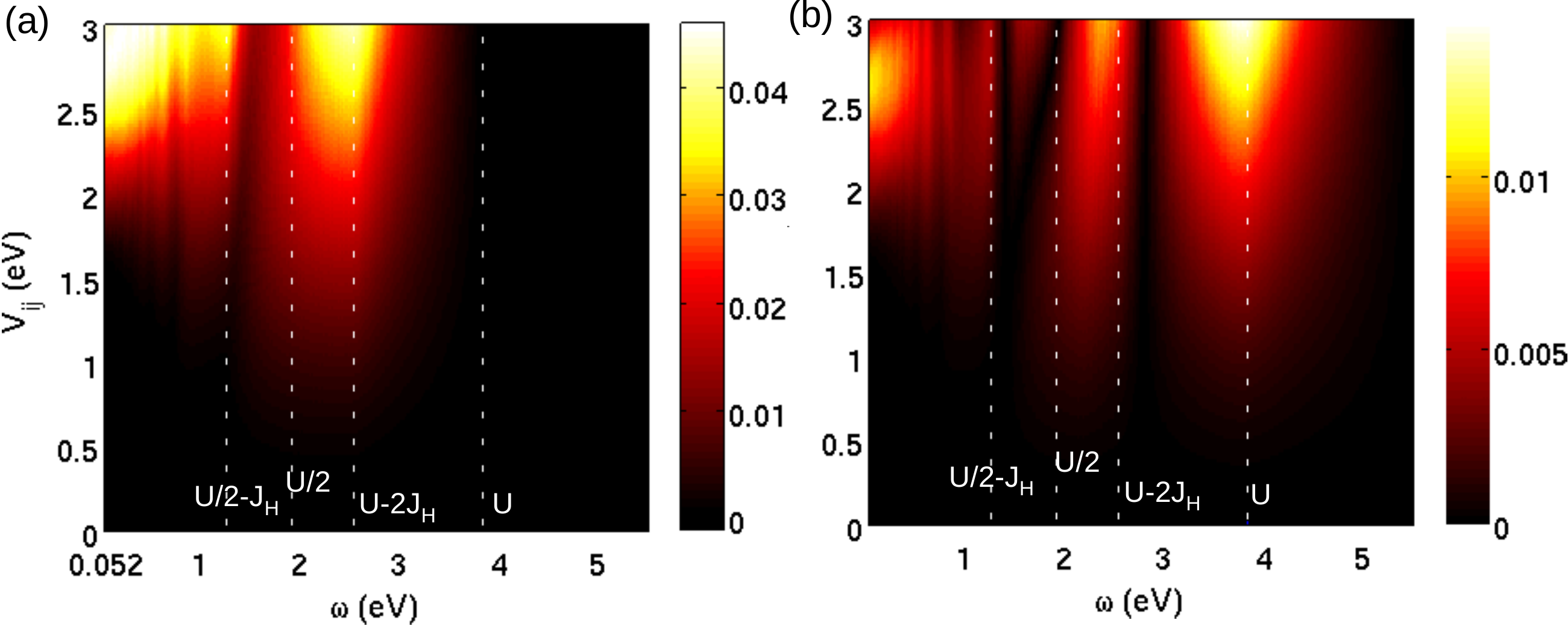}
\caption{The imaginary part of the Floquet spin-orbital Hamiltonian projected to the bond $\langle ij \rangle$,  
(a) for YTiO$_3$ and  (b) LaTiO$_3$.}
\label{fig:heating}
\end{figure}

Including the hopping within the DH subspace  smears out the
sharp patterns of the effective exchange interactions shown in
Fig.~(\ref{fig:floquet-exchange})(a)-(b), and introduces
imaginary parts to the Floquet spin-orbital Hamiltonians when $\omega$
is in resonance with the excitation bands.  In
Fig.~\ref{fig:floquet-exchange}(c) and (d) we plot the real parts of
the in-plane effective exchange interactions for YTiO$_3$ and
LaTiO$_3$, taking into account the bandwidth of the excitation spectra
due to the DH hopping. The interesting features in
Fig.~\ref{fig:floquet-exchange}(a)-(b) are mostly preserved in
Fig.~(\ref{fig:floquet-exchange})(c)-(d), except that the sign flip of
the exchange interaction in LaTiO$_3$
(Fig.~(\ref{fig:floquet-exchange})(c)) has been completely smeared out
by DH hopping. We also plot the imaginary part of
$H_{\textrm{F}}^{\textrm{so}}$ (Eq.~(\ref{eq:h-floquet-t2g}))
projected to the in-plane bond $\langle ij\rangle$ for both YTiO$_3$
((a)) and LaTiO$_3$ ((b)). We have assumed that the spin order is
ferromagnetic (antiferromagnetic) in YTiO$_3$ (LaTiO$_3$).
Clearly $\textrm{Im}[H_{\textrm{F}}^{\textrm{so}}]$ has a broad peak
centered at $U$ ($U-2J_{\textrm{H}}$) for the antiferromagnet
(ferromagnet). Moreover in LaTiO$_3$
$\textrm{Im}[H_{\textrm{F}}^{\textrm{so}}]$ is significant between
$U/2$ and $U-2J_{\textrm{H}}$, which invalidates the description of
the system by an effective spin(-orbital) Hamiltonian in this
frequency
regime. 

To summarize, we have derived the Floquet spin-orbital model for
multi-orbital Mott insulators using time-dependent perturbation
theory, taking into account the effects of the bandwidth of the DH
excitations.  We have applied our formalism to orthorhombic perovskite
titanates YTiO$_3$ and LaTiO$_3$ based on first-principles
calculations.  In certain frequency regimes the effective spin
exchange interactions are highly tunable by the laser field, and
exhibit robust features which may be experimentally measurable.  The
formalism and methodology presented in this paper can be directly
applied to Slater/Mott insulators with any kind of ordered
spin-orbital ground state, which may stimulate further exploration of
Floquet engineering of magnetism in strongly correlated
transition-metal oxides.

\textit{Acknowledgements: } This research was supported by the NSF
materials theory program through grant DMR1506119 (LB, KH) and the Army
Research Office MURI grant ARO W911NF-16-1-0361, Floquet engineering and
metastable states (JL).


\bibliography{yto}

\begin{thebibliography}{51}
\bibitem[{com()}]{comment_nphoton}
\bibinfo{note}{The important photon absorption/emission processes are dominated
  by those with $n$ being on the order of unity. Because when $n$ is large, the
  corresponding weight factor $\mathcal{J}_{n}(\mu_{ij})^2$ becomes
  negligible.}
\end{thebibliography}
\end{document}